\documentstyle[preprint,epsfig,aps]{revtex}
\tightenlines
\newcommand{\bs}{\bf}
\begin{document}
\draft

\title{Pion cloud and the $Q^2-$dependence of\\
 $\gamma^* N \leftrightarrow \Delta$ transition form factors}
\author{S. S. Kamalov\cite{Sabit} and Shin Nan Yang}
\address{Department of Physics, National Taiwan University, Taipei 10617,
Taiwan}

\date{\today}
\maketitle

\begin{abstract}
Recent experiment indicates that the behavior exhibited by the ratios
$E_{1^+}/M_{1^+}$ and $S_{1^+}/M_{1^+}$ of the
$\gamma^* N \leftrightarrow \Delta$ transition
remain small and {\it negative} for  $Q^2 \le 4.0\,\,GeV^2$. It implies
that the perturbative QCD is still not applicable at these momentum
transfers. We show that these data can be explained in a
dynamical model for electromagnetic production of pions, together with a
simple scaling assumption for the bare $\gamma^* N \Delta$ form factors.
Within our model we find that the bare $\Delta$ is almost spherical and the
electric E2 and Coulomb C2 quardrupole excitations of the physical $\Delta$
are nearly saturated by pion cloud contribution in $Q^2 \le 4.0\,\,GeV^2$.
\end{abstract}
%
%

 It has been well recognized that the study of the excitations of the
hadrons can shed light on the nonperturbative aspects of QCD. One
case which has recently been under intensive study is the
electromagnetic excitation of the $\Delta(1232)$ resonance. At low
four-momentum transfer squared $Q^2$,  the interest is motivated
by the possibility of observing a $D-$state in the $\Delta$
\cite{Beck97,Hsiao98,Davidson91}. The existence of a $D-$state in
the $\Delta$ has the consequence that the $\Delta$ is deformed and
the photon can excite a nucleon through electric $E2$ and Coulomb
$C2$ quardrupole transitions. In a symmetric SU(6) quark model,
the electromagnetic excitation of the $\Delta$ could proceed only
via $M1$ transition. In pion electroproduction, $E2$ and $C2$
excitations would give rise to nonvanishing $E_{1+}^{(3/2)}$ and
$S_{1+}^{(3/2)}$ multipole amplitudes. Recent experiments give,
near $Q^2 =0$, $R_{EM} = E_{1+}^{(3/2)}/M_{1+}^{(3/2)} \sim -0.03$
\cite{Beck97} a clear indication of $\Delta$ deformation.

At sufficiently large $Q^2$, the perturbative QCD (pQCD) is expected
to work. It predicts that only helicity-conserving amplitudes
contribute at high $Q^2$ \cite{Brodsky81}, leading to
 $R_{EM} = E_{1+}^{(3/2)}/M_{1+}^{(3/2)} \rightarrow 1$ and
 $R_{SM} = S_{1+}^{(3/2)}/M_{1+}^{(3/2)} \rightarrow const$.
This behavior in the perturbative domain is very different from
that in the nonperturbative one. It is an intriguing question to
find the region of $Q^2$ which signals the onset of the pQCD.

In a recent measurement \cite{Frolov99}, the electromagnetic
excitation of the $\Delta$ was studied at $Q^2=2.8$ and $4.0
\,\,GeV^2$ via the reaction $p(e,e'p)\pi^0$. The extracted ratios
$R_{EM}$ and $R_{SM}$ remain small and {\it negative}. This
disagrees with the previous analysis \cite{Burkert95} of the
earlier DESY data \cite{Haidan79} which gave small but {\it
positive} $R_{EM}$ and $R_{SM}$ at $Q^2 = 3.2 \,\,GeV^2$, though
both analyses indicate that pQCD is still not applicable in this
region of $Q^2$. In this Letter, we want to show that the recent
data of Ref. \cite{Frolov99} can be understood from the dominance
of the pion cloud contribution at low $Q^2$, in both
$E_{1+}^{(3/2)}$ and $S_{1+}^{(3/2)}$, as predicted by a dynamical
model \cite{Yang85,Tanabe85} for electromagnetic production of
pion, together with a simple scaling assumption for the bare
$\gamma^* N\Delta$ form factors.

The main feature of the dynamical approach to the pion photo- and
electroproduction \cite{Yang85,Tanabe85} is that the unitarity is
built in by explicitly including the final state $\pi N$
interaction in the theory, namely, t-matrix is expressed as
\begin{eqnarray}
t_{\gamma\pi}(E)=v_{\gamma\pi}+v_{\gamma\pi}\,g_0(E)\,t_{\pi
N}(E)\,, \label{eq:tgamapi}
\end{eqnarray}
where $v_{\gamma\pi}$ is the transition potential operator for
$\gamma^*N \rightarrow \pi N$ and, $t_{\pi N}$ and $g_0$ denote
the $\pi N$ t-matrix and free propagator, respectively, with $E$
the total energy in the CM frame.

In the (3,3) channel where $\Delta$ excitation plays an important role,
the transition potential $v_{\gamma\pi}$ consists of two terms
\begin{eqnarray}
v_{\gamma\pi}(E)=v_{\gamma\pi}^B + v_{\gamma\pi}^{\Delta}(E)\,,
\label{eq:tranpot}
\end{eqnarray}
where $v_{\gamma\pi}^B$ is the background transition potential
which includes Born terms and vector mesons exchange
contributions, as described in Ref. \cite{UIM99}.
The second term of Eq. (\ref{eq:tranpot})
corresponds to the contribution of bare $\Delta$,
namely,  $\gamma^* N \rightarrow \Delta \rightarrow \pi N$.

In accordance with Ref. \cite{Aron93}, we decompose Eq. (\ref{eq:tgamapi})
in the following way, as shown in Fig. 1,
\begin{eqnarray}
t_{\gamma\pi}= t_{\gamma\pi}^B + t_{\gamma\pi}^{\Delta}\,,
\label{eq:decomp}
\end{eqnarray}
where
\begin{eqnarray}
t_{\gamma\pi}^B(E)=v_{\gamma\pi}^B+v_{\gamma\pi}^B\,g_0(E)\,t_{\pi
N}(E)\,,\\ \label{eq:decompa}
t_{\gamma\pi}^\Delta(E)=v_{\gamma\pi}^\Delta+v_{\gamma\pi}^\Delta\,
g_0(E)\,t_{\pi N}(E)\,.
\label{eq:decompb}
\end{eqnarray}
The advantage of such a decomposition (\ref{eq:decomp}) is that
all the processes which start with the electromagnetic excitation
of the bare $\Delta$  are summed up in $t_{\gamma\pi}^\Delta$. The
solid blob for both the intermediate $\Delta$ states and the $\pi
N \Delta$ vertex means that they are dressed \cite{Hsiao98,Tanabe85}. Eq.
(\ref{eq:decomp}) provides us with a prescription to extract
information concerning bare $\Delta$ excitation.


For physical multipole amplitude in channel $\alpha$, multipole
decomposition of $t_{\gamma\pi}$ gives \cite{Yang85}
\begin{eqnarray}
& & t^{(\alpha)}_{\gamma\pi}(q_{E}, k_E;E+i\varepsilon) =
e^{i\delta_{\alpha}}\cos\delta_{\alpha}\,[
v^{(\alpha)}_{\gamma\pi}(q_{E}, k_E)+ \nonumber\\ & + &
P\int^{\infty}_{0}d q'\,\frac{q'^2 R^{(\alpha)}_{\pi N}
(q_{E},q';E)\,v^{(\alpha)}_{\gamma\pi}(q',k_E)}{E-E_{\pi
N}(q')}\,]\,, \label{eq:mult}
\end{eqnarray}
where $\delta_{\alpha}$, $ R_{\pi N}^{(\alpha)}$, $E_{\pi N}(q)$
and $P$ denote the $\pi N$ phase shift, reaction matrix in channel
$\alpha$, total CM energy of momentum $q$, and principal value
integral, respectively; $k_E=\mid{\bf k}\mid$ is the photon
momentum and $q_E$ the pion on-shell momentum. Eq. (\ref{eq:mult})
manifestly satisfies Watson theorem and shows that $\gamma\pi$
multipoles depend on the half-off-shell behavior of the
pion-nucleon interaction. To make principal value integration
associated with $v_{\gamma\pi}^B$ convergent, we introduce an
off-shell dipole form factor, which characterizes the finite range
aspects of the potential. The cut-off parameter $\Lambda$ is
determined by requiring that it provides the best fit to the
$M_{1+}^{(3/2)}$, which turns out to be $\Lambda$=440 MeV.

 Note that due to the off-shell rescattering effects in the principal
value integral of Eq. (\ref{eq:mult}), gauge invarinace is violated.
In the present model for the pion electroproduction
we restore gauge invariance by the following substitution
\begin{eqnarray}
J_{\mu}^B \rightarrow J_{\mu}^B  - k_{\mu}\frac{k\cdot
J^B}{k^2}\,, \label{eq:gauginv}
\end{eqnarray}
where $J_{\mu}^B$ is the electromagnetic current corresponding to
the background  contribution $v_{\gamma\pi}^B$.

We evaluate $t_{\gamma\pi}^B$ with $t_{\pi N}$ matrix elements
obtained in a meson-exchange model \cite{Hung94}. From the
structure of $t_{\gamma\pi}^{\Delta}$ as depicted in Fig. 1, we
can describe its energy dependence of the corresponding multipole
amplitudes $A^{\Delta}$ with a Breit-Wigner form, as
was done in the isobar model of Ref. \cite{UIM99},
\begin{equation}
A^{\Delta}(W,Q^2)\,=\,{\bar{\cal A}}^{\Delta}(Q^2)\, \frac{
f_{\gamma \Delta}\,\Gamma_{\Delta}\,M_{\Delta}\,f_{\pi \Delta} }
 {M_{\Delta}^2-W^2-iM_{\Delta}\Gamma_{\Delta}}\,e^{i\phi}\,,
\label{eq:BW}
\end{equation}
where $f_{\pi \Delta}(W)$ is the usual Breit-Wigner factor
describing the decay of the $\Delta$ resonance with total width
$\Gamma_{\Delta}(W)$ and physical mass $M_{\Delta}$=1232 MeV. The
$W$ dependence of the $\gamma N \Delta$ vertex is given in
$f_{\gamma \Delta}(W)$ with normalization $f_{\gamma
\Delta}(M_{\Delta})=1$. The expressions for $f_{\gamma \Delta}, \,
f_{\pi \Delta}$ and $\Gamma_{\Delta}$ are taken from Ref.
\cite{UIM99}. The phase $\phi(W,Q^2)$ in Eq. (\ref{eq:BW}) is to
adjust the phase of $A^{\Delta}$ to be equal to the corresponding
pion-nucleon scattering phase $\delta_{33}$. At the resonance
$\phi(M_{\Delta},Q^2)=0$ and it does not affect the $Q^2$
dependence of the electromagnetic vertex.

The main parameters in the bare $\gamma^* N \Delta$ vertex are the
${\bar{\cal A}}^{\Delta}$'s in Eq. (\ref{eq:BW}). For the magnetic
dipole ${\bar{\cal M}}^{\Delta}$ and electric quadrupole
${\bar{\cal E}}^{\Delta}$ transitions they are related to the
conventional electromagnetic helicity amplitudes $A^\Delta_{1/2}$
and $A^\Delta_{3/2}$ by
\begin{eqnarray}
& & {\bar{\cal M}^\Delta}=-\frac{1}{2}(A^\Delta_{1/2} + \sqrt{3}
A^\Delta_{3/2})\,,\nonumber\\ & & {\bar{\cal
E}^\Delta}=\frac{1}{2}(-A^\Delta_{1/2} + \frac{1}{\sqrt{3}}
A^\Delta_{3/2})\,.
\end{eqnarray}
For the real photons, they are equal to the standard $M1$ and $E2$
amplitudes of the $\gamma N\rightarrow\Delta$ transition as
defined by the Particle Data Groups.


In the present work, we parametrize the $Q^2$ dependence of the
dominant ${\bar{\cal M}}^\Delta$ amplitude by
\begin{eqnarray}
{\bar{\cal M}}^\Delta(Q^2)={\bar{\cal M}}(0)
\frac{\mid {\bs k}\mid}{k_{\Delta}}\,(1+\beta Q^2)\,
e^{-\gamma Q^2}\, G_D(Q^2)\,,
\label{eq:q2ansatz}
\end{eqnarray}
where $G_D$ is the nucleon dipole form factor. The parameters
$\beta$ and $\gamma$ will be determined later. For the small
${\bar{\cal E}}^\Delta$ and ${\bar{\cal S}}^\Delta$ amplitudes,
following Refs. \cite{UIM99,Laget88}, we assume that they have the
same $Q^2$ dependence as ${\bar{\cal M}}^\Delta$. This is
motivated by the scaling law which has been observed for the
nucleon form factors. It is plausible if a bare $\Delta$ is
pictured as simply flipping one of the quark spins in the nucleon.
It is also known that ${\bar{\cal E}}^\Delta(0)= {\bar{\cal
S}}^\Delta(0)$ \cite{Laget88}.

 To proceed, we first consider $M_{1+}^{(3/2)}$ and
$E_{1+}^{(3/2)}$ multipoles at $Q^2=0$. By combining the
contributions of $t_{\gamma\pi}^B$ and $t_{\gamma\pi}^\Delta$ and
using the bare amplitudes ${\bar{\cal M}}^\Delta (0)$ and
${\bar{\cal E}}^\Delta (0)$ of Eq. (\ref{eq:BW}) as free
parameters, results of our best fit to the multipoles obtained in
the recent analyses of Mainz~\cite{HDT} and VPI group~\cite{VPI97}
are shown in Fig. 2 by solid curves. The dashed curves denote the
contribution from $t_{\gamma\pi}^B$ only. The dotted curves
represented the K-matrix approximation to $t_{\gamma\pi}^B$,
namely, without the principal value integral term included.

The numerical values obtained for ${\bar{\cal M}}^\Delta$ and
${\bar{\cal E}}^\Delta$ and the helicity amplitudes, at $Q^2 =0$,
are given in Table 1 along with the corresponding  "dressed"
values. At the resonance position $t_{\gamma\pi}^B$ vanishes
within K-matrix approximation and only principal value integral
term survives. The latter corresponds to the contribution where
$\Delta$ is excited by the pion produced via $v_{\gamma\pi}^B$.
Consequently the addition of this contribution to
$t_{\gamma\pi}^{\Delta}$ can be considered as a dressing of the
$\gamma N \Delta$ vertex. The dressed helicity amplitudes obtained
in this way are in very good agreements with the results of
Ref.\cite{UIM99} and with PDG values.

One notices that the bare values for the helicity amplitudes
determined above, which amount to only about $60\%$ of the
corresponding dressed values, are close to the predictions of the
constituent quark model (CQM), as was pointed out by Sato and Lee
\cite{Sato96}. The large reduction of the helicity amplitudes from
the dressed to the bares ones result from the fact that the
principal value integral part of Eq. (\ref{eq:mult}), which
represents the effects of the off-shell pion rescattering,
contributes approximately for half of the $M_{1+}$ as indicated by
the dashed curves in Fig. 2.



For the standard Sach-type form factor
$G_M^{\Delta}(0)$~\cite{Jones} our bare and dressed values are
$1.65\pm 0.02$ and $3.06\pm 0.02$, respectively. On the other
hand, results of CQM calculations lie in the range 1.4--2.2
\cite{CQM}. From this result we conclude that pion rescattering is
the main mechanism responsible for the longstanding discrepancy in
the description of the magnetic $\gamma^*N\rightarrow\Delta$
transition within CQM. For $E_{1+}^{(3/2)}$, the dominance of
background and pion rescattering contributions further leads to a
very small bare value for electric  transition.

We now turn to the $Q^2$ evolution of the multipoles in the (3,3) channel.
With the  parametrization of (\ref{eq:q2ansatz}), we fit the
recent experimental data \cite{Frolov99} as well as old one quoted in
Ref. \cite{UIM99} on the $Q^2$ dependence of  
$M_{1+}^{(3/2)}$  multipole or equivalently, the $G_M^*$ form factor
defined as \cite{UIM99}
\begin{equation}
M_{1+}^{3/2}(M_\Delta,Q^2) = \frac{\mid {\bf k}\mid}{m_N}
\sqrt{\frac {3\alpha}{8\Gamma_{exp}q_\Delta}}G_M^*(Q^2)
\end{equation}
with $\alpha =1/137$, $\Gamma_{exp}$=115 MeV, $q_{\Delta}$ is the
pion momentum at the resonance energy and $m_N$ is the nucleon
mass. Our result is shown in Fig. 3. Note that for the $G_M^*$
form factor we use the "Ash" definition~\cite{Ash67}. It differs
from the definition used in Refs{\cite{Frolov99,Jones} by factor
$[1+Q^2/(m_N+M_{\Delta})^2]^{1/2}$. At $Q^2=0$ we have
$G_M^*(0)=G_M^{\Delta}(0)$. The obtained values for the $\beta$
and $\gamma$ parameters of Eq. (\ref{eq:q2ansatz}) are:
 $\beta=0.44 \,\,GeV^{-2}$ and  $\gamma=0.38\,\, GeV^{-2}$. Here the
dashed curves correspond to contribution from the  bare $\Delta$, i.e.,
$t_{\gamma\pi}^\Delta$ of Eq. (\ref{eq:decompb}).


With the scaling assumption, i.e., both ${\bar{\cal E}}^\Delta$
and ${\bar{\cal S}}^\Delta$ have the same $Q^2$ dependence as
${\bar{\cal M}}^\Delta$ as given in Eq. (\ref{eq:q2ansatz})
and ${\bar{\cal E}}^\Delta(0)$ determined above, together with  the
relation ${\bar{\cal E}}^\Delta(0)={\bar{\cal S}}^\Delta(0)$,
the $Q^2$ dependence for the ratios
$R_{EM}=E_{1+}^{(3/2)}/M_{1+}^{(3/2)}$  and
$R_{SM}=S_{1+}^{(3/2)}/M_{1+}^{(3/2)}$ can be evaluated. The
results are  shown in Fig. 4. It is seen that they are in good
agreement with the results of the model independent analysis of
Ref .\cite{Frolov99} up to $Q^2$ as high as $4.0\,\,GeV^2$. Note
that since the bare values for the electric and Coulomb
excitations are small, the absolute values and shape of these
ratios are determined, to a large extent, by the pion rescattering
contribution. The bare $\Delta$ excitation contributes mostly to
the $M_{1+}^{(3/2)}$ multipole.


In summary, we calculate the $Q^2$ dependence of the ratios
$E_{1+}/M_{1+}$ and $S_{1+}/M_{1+}$ in the
$\gamma^*N\rightarrow\Delta$ transition, with the use of a
dynamical model for electromagnetic production of pions and a
simple scaling assumption for the bare $\gamma^* N\rightarrow
\Delta$ transition form factors. We find that both ratios
$E_{1+}/M_{1+}$ and $S_{1+}/M_{1+}$ remain small and {\it negative} for
$Q^2 \le 4.0\,\,GeV^2$. Our results agree well with the recent
measurement of  Ref. \cite{Frolov99}, but deviate strongly
from the predictions of pQCD. Our results indicate that the bare
$\Delta$ is almost spherical and hence very difficult to be
directly excited via electric E2 and Coulomb C2 quardrupole
excitations. The experimentally observed $E_{1+}^{(3/2)}$ and
$S_{1+}^{(3/2)}$ multipoles are, to a very large extent, saturated
by the contribution from pion cloud, i.e., pion rescattering
effects. It remains an intriguing question, both theoretically and
experimentally, to find the region of $Q^2$ which will signal the
onset of pQCD.

\acknowledgements We would like to thank Guan-Yu Chen for the
installation of the meson exchange pion-nucleon scattering code and
Shian-Shyong Hsiao for the useful discussions. SSK would like to thank the
Department of Physics at the National Taiwan University  for warm
hospitality and National Science Council of the Republic of China
for financial support. This work is supported in part by the
NSC/ROC under the grant no. NSC 88-2112-M002-015.

\newpage


\vspace{-0.5cm}

\hspace{-0.1in}\begin{minipage}{3.4in}
\begin{table}[htbp]
\begin{tabular}{|c|ccc|}
             Amplitudes            &  "bare"         & "dressed"    & PDG \\
 \hline ${\bar{\cal M}}^{\Delta}$  & $ 158\pm 2 $    & $289\pm 2 $  & $293\pm 8$ \\
        ${\bar{\cal E}}^{\Delta}$  & $ 0.4\pm 0.3 $  & $-7\pm 0.4 $ & $-4.5 \pm 4.2 $ \\
        $ A_{1/2}^{\Delta}$        & $ -80 \pm 2 $   & $ -134\pm 2$ & $-140\pm 5 $\\
        $ A_{3/2}^{\Delta}$        & $ -136 \pm 3 $  & $ -256\pm 2$ & $-258\pm 6$\\
\end{tabular}

\vspace{0.2cm}

\caption {Comparison of the "bare" and "dressed" values for the
amplitudes ${\bar{\cal A}}^{\Delta},\, A_{1/2}^{\Delta}$ and
$A_{3/2}^{\Delta}$ (in $10^{-3}\,GeV^{-1/2}$).}
\end{table}
\end{minipage}

\begin{figure}[tbp]
\begin{center}
\epsfig{file=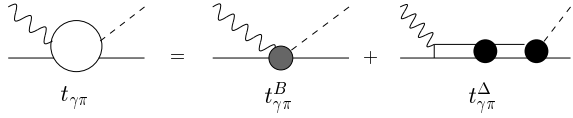,width=3.3in}
\end{center}
\caption{ Graphical representation of the pion electroproduction
$t$ matrix.}\label{figks1}
\end{figure}

\begin{figure}[tbp]
\begin{center}
\epsfig{file=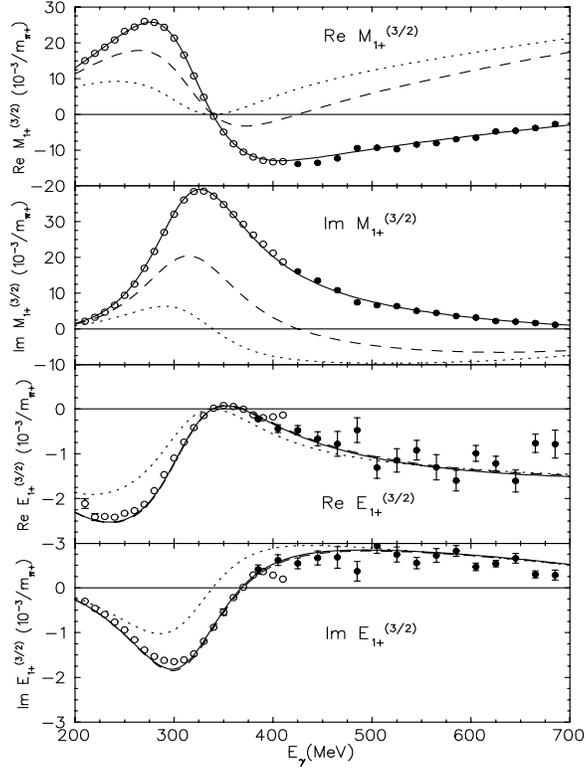,width=3.0in,height=4.0in}
\end{center}
\caption{ Real and imaginary parts of the $M_{1+}^{(3/2)}$ and
$E_{1+}^{(3/2)}$ multipoles. Dotted and dashed curves are the
results for the $t_{\gamma\pi}^B$ obtained without and with
principal value integral contribution in Eq. (6), respectively.
Solid curves are the full results with bare $\Delta$ excitation.
For the $E_{1+}$ dashed and solid curves are practically the same
due to the small value of the bare ${\bar{\cal E}}^{\Delta}$. The
open and full circles are the results from the Mainz dispersion
relation analysis~\protect\cite{HDT} and  from the VPI
analysis~\protect\cite{VPI97}. }\label{figks2}
\end{figure}

\begin{figure}[t]
\begin{center}
\epsfig{file=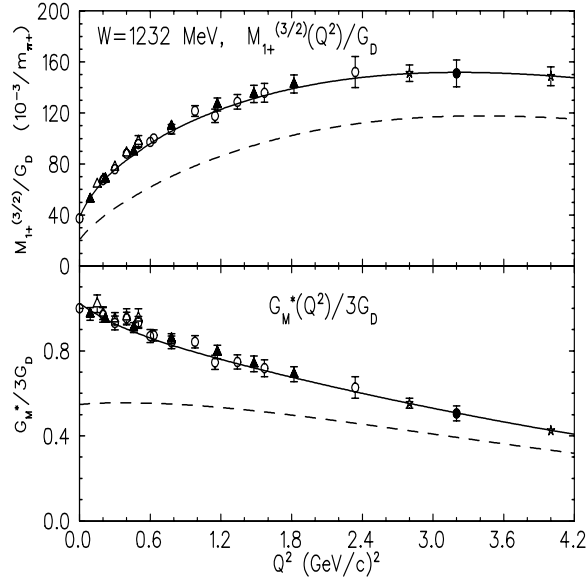,width=3.0in,height=3in}
\end{center}
\caption{ The $Q^2$ dependence of  $Im\,M_{1+}^{(3/2)}$ at
$W$=1232 MeV and corresponding $G_M^*$ form factor. The full and
dashed curves are the results for the "dressed" and "bare" $\gamma
N \Delta$ vertexes, respectively. Experimental data are from
Ref.\protect\cite{EXPM}. The new data at $Q^2$=2.8 and 4.0
$(GeV/c)^2$ are from Ref.\protect\cite{Frolov99}. }\label{figks3}
\end{figure}

\begin{figure}[t]
\begin{center}
\epsfig{file=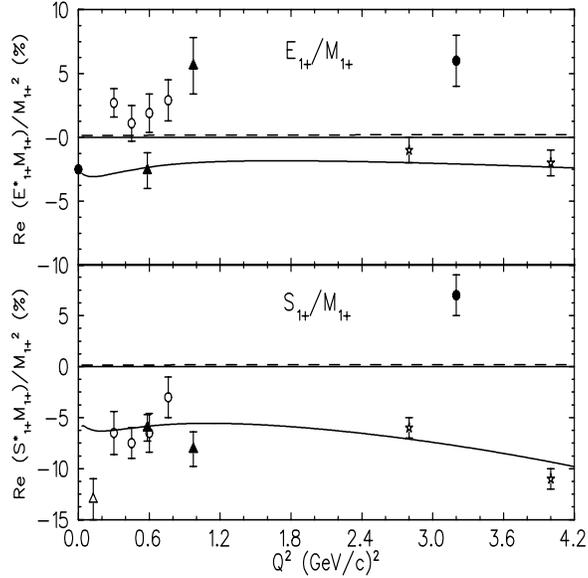,width=3.0in,height=3in}
\end{center}
\caption{ The $Q^2$ dependence of the ratios
$E_{1+}^{(3/2)}/M_{1+}^{(3/2)}$ and
$S_{1+}^{(3/2)}/M_{1+}^{(3/2)}$ at $W=1232$ MeV. Notations for the
curves are the same as in Fig. 3. Experimental data are from
Ref.\protect\cite{EXPR}. The new data at $Q^2$=2.8 and 4.0
$(GeV/c)^2$ are the results of the model independent multiple
analysis of Ref.\protect\cite{Frolov99}. } \label{figks4}
\end{figure}

\end{document}